\documentclass[journal]{IEEEtran}
\usepackage{graphicx}
\usepackage{textcomp}

\begin{document}

\title{Latest developments on the highly granular Silicon-Tungsten Electromagnetic Calorimeter technological prototype for the International Large Detector.}

\author{A. Irles\thanks{Laboratoire de l'Acc\'el\'erateur Lin\'eaire, Centre Scientifique d'Orsay, Universit\'e de Paris-Sud XI, CNRS/IN2P3, F-91898 Orsay Cedex, France (e-mail: \texttt{irles{@}lal.in2p3.fr})} on behalf of the CALICE Collaboration.}


\maketitle

\begin{abstract}
  High precision physics at future colliders requires unprecedented
  highly granular calorimeters for the application of the Particle
  Flow (PF) algorithm. 
  The physical proof of concept was given in the previous campaign
  of beam tests of physic prototypes within the CALICE collaboration.
  We present here the latest beam and laboratory test results and
  R\&D developments for the Silicon-Tungsten Electromagnetic Calorimeter
  technological prototype with fully embedded very front-end (VFE)
  electronics for the International Large Detector at the
  International Linear Collider project. 
\end{abstract}


\IEEEpeerreviewmaketitle

\section{Introduction}

\IEEEPARstart{F}{uture} accelerator based particle physics experiments
require very precise and detailed reconstruction of the final states produced
in the beam collisions. A particular example is the next generation of $e^{+}e^{-}$
linear colliders such the ILC\cite{ilctdr}.
This project will provide collisions of polarized beams with centre-of-mass energies ($c.m.e$) of 250 GeV - 1 TeV.
These collisions will be studied by two multipurpose detectors:
the International Large Detector (ILD) and the Silicon Detector (SiD)\cite{Behnke:2013lya}.
Another example of an $e^{+}e^{-}$ collider project is the Compact Linear Collider (CLIC)
project\cite{clictdr} which will produce collisions with $c.m.e$ of 380 GeV - 3 TeV
with a detector featuring similar design than the ILD and SiD.
Both projects will study with unprecedented precision the final states with heavy bosons (W, Z  and H) and heavy quarks ($c$, $b$ and $t$).

To meet the required precision levels, these detectors will be based on the Particle Flow (PF) techniques\cite{Brient:2002gh,Morgunov:2004ed}.
These techniques consist of choosing the best information available
to measure the energy of the final state objects (i.e. measuring the charged particles momentum at tracking devices better than in the calorimeters).
Therefore, PF techniques rely on single particle separation.
For this purpose, PF algorithms require highly granular and compact calorimeter systems featuring minimum dead material.
The R\&D of highly granular calorimeters for future linear colliders is conducted within the CALICE collaboration\cite{calice}.
For further about PF and CALICE R\&D we refer the reader to reference \cite{Sefkow:2015hna} and references therein. 

Due to the PF requirements, the calorimeters at ILD will be placed inside the magnetic coil
providing magnetic fields of 3.5 T. The baseline design of the ILD ECAL consists in a detector (in the barrel region) of 24 $X_{0}$ of thickness.
The silicon-tungsten electromagnetic calorimeter, SiW-ECAL, design and R\&D conducted by CALICE is oriented at the baseline design of the ILD ECAL.
It has silicon (Si) as active material and tungsten (W) as absorber material.
The combination of Si and W for the construction of the layers allows
for building a very compact calorimeter with compact active layers and small cell size (high granularity) in the transverse and longitudinal planes.
It will consist of an alveolar structure of carbon fibre into which slabs made of tungsten
plates and the active sensors will be inserted. The very-front-end (VFE) electronics will be
embedeed in the slabs. The silicon sensors will be segmented
in squared cells of 5x5 mm: a total of $\sim 100$ million channels will constitute the ECAL for ILD.
To reduce overall power consumption, the SiW-ECAL will exploit the special bunch structure
forseen for the ILC: the $e^{+}e^{-}$ bunchs trains will arrive within
acquisition windows of $\sim$ 1-2 ms width separated by $\sim$ 200 ms. During the idle time, the bias currents of the electronics will be shut down.
This technique is usually denominated power pulsing.

\section{The CALICE silicon tungsten electromagnetic calorimeter engineering prototype.}

The first SiW-ECAL prototype was the so called SiW-ECAL physics prototype.
It was successfully tested at DESY, FNAL and CERN running together with another CALICE prototype,
the analogue hadronic calorimeter AHCAL delivering the proof of concept of PF calorimetry.
For the physics prototype, the VFE was placed outside the active area with no particular constraints in power consumption.
It consisted of 30 layers of Si as active material alternated with tungsten plates as absorber material.
The active layers were made of a matrix of 3x3 Si wafers. Each of these wafers was segmented in matrices of
6x6 squared pixels of 1x1 $cm^{2}$.
The prototype was divided in 3 modules of 10 layers with different W depth per layer in each of these modules
(0.4, 1.6 and 2.4 $X_{0}$) making a total of 24 $X_{0}$ which equivales to $\sim 1~\lambda_{I}$ (interaction length).
Published results proving the good performance of the technology and the PF can be found in references \cite{Adloff:2011ha,Anduze:2008hq,Adloff:2008aa,Adloff:2010xj,CALICE:2011aa} and \cite{Bilki:2014uep} 

Current prototype, the technological prototype, addresses the main technological challenges: compactness,
power consumption reduction through power pulsing and VFE inside the detector close to real ILD conditions.
It will also provide data to deeply study the PF and provide input to tune the Monte Carlo programs.

\subsection{Slabs and Active Signal Units.}

\begin{figure}[!t]
\centering
\includegraphics[width=3.0in]{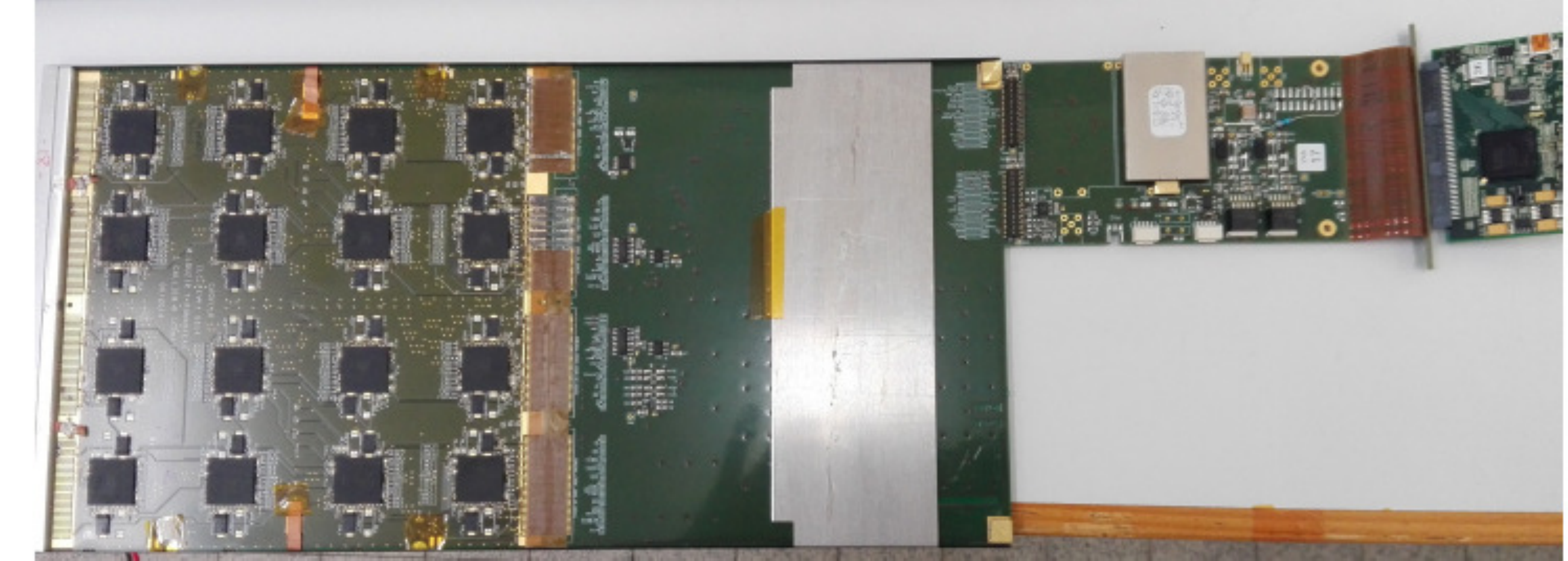}
\caption{Open single SLAB with FEV11 ASU, 16 SKIROC, interface card and DIF visibles. The silicon sensors are glued to the PCB in the other side.}
\label{shortslab}
\end{figure}

\begin{figure}[!t]
  \centering
  \begin{tabular}{l}
    \includegraphics[width=3.0in]{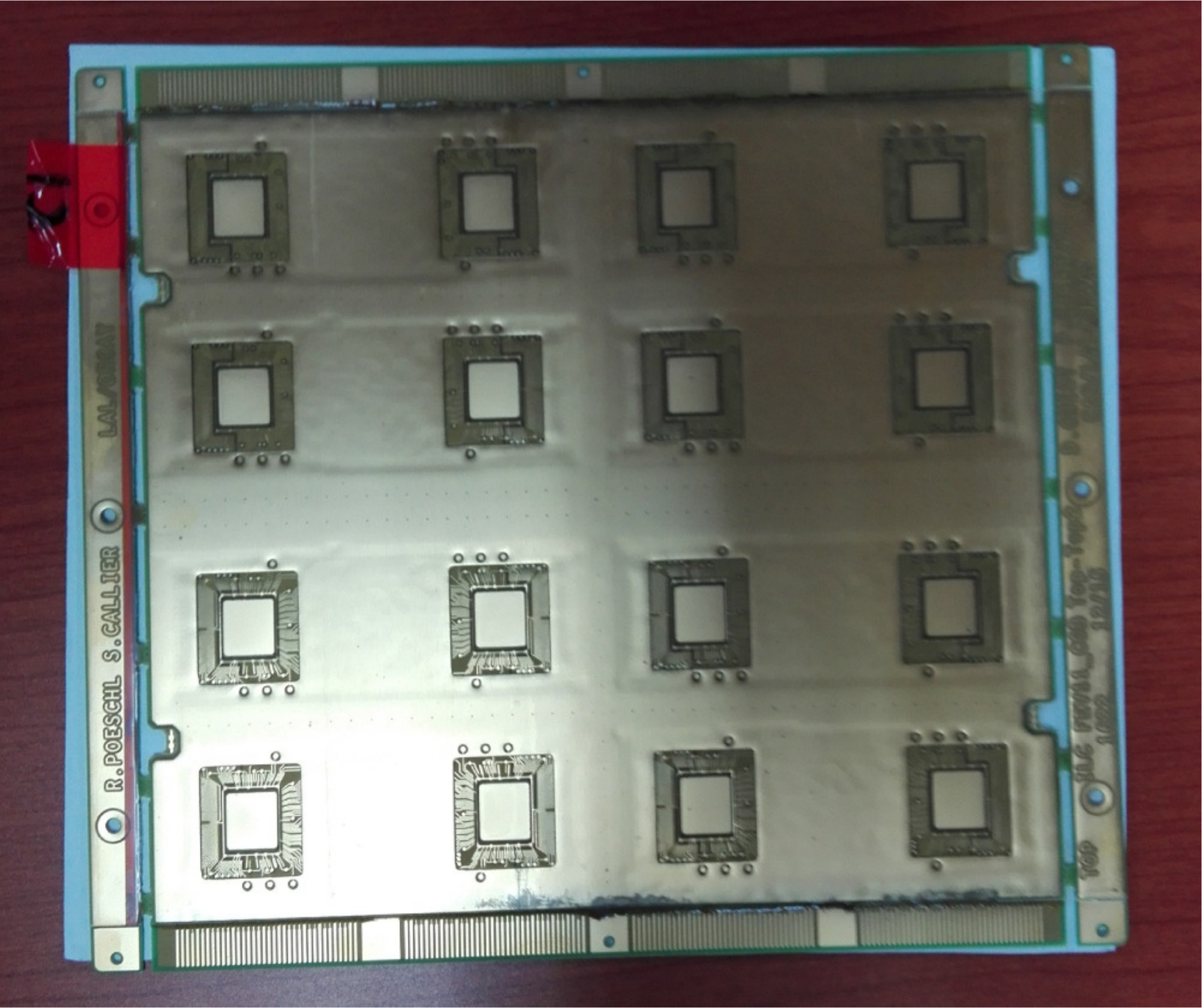}\\
    \includegraphics[width=3.0in]{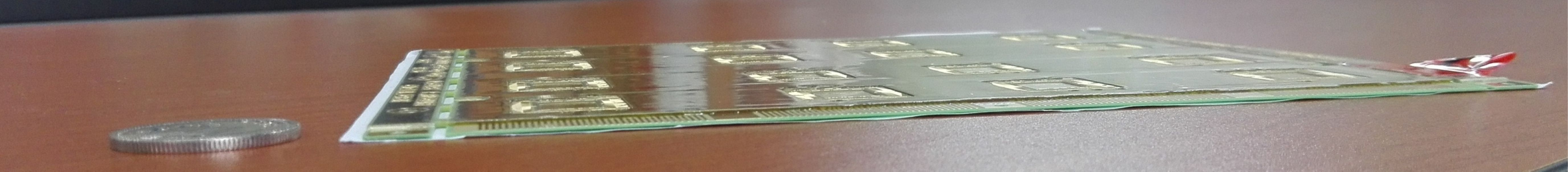}
  \end{tabular}
\caption{Photograph of the FEV11\_COB. The first figure correspond to a picture taken from the top. The second one is a lateral picture.}
\label{fevcob}
\end{figure}

The entity
of sensors, thin PCB (printed circuit boards) and ASICs (application-specific integrated circuits) is called Active Signal Units or ASU.
An individual ASU has a lateral dimension of 18x18 cm$^{2}$ and has glued onto it 4 silicon wafers (currently with a thickness of 320 $\mu$m).
The ASUs are equipped
further with 16 ASIC for the read out and features 1024 square pads (64 per ASIC) of 5x5 mm.
The readout layers of the SiW-ECAL consist of a chain of ASUs and an interface card
to a data acquisition system (DAQ) at the beginning of the layer.
This interface card also carries services as power connectors,
test output pins, connectors for signal injection, etc. 
Currently, the technological prototype layers are built with version of PCB called FEV11 with 16 SKIROC (see section \ref{sec:DAQ})
 in BGA packages mounted on top of it.
These PCBs still don't meet the requirements
for the ILD in terms of thickness (1.2 mm). The FEV11 thickness is 1.6 alone and 2.7 mm including the ASICs.
Figure \ref{shortslab} shows a picture of a full equipped short slab with FEV11 ASU.
There are ongoing R\&D activities in an alternative PCB design in which the ASICs
will be directly placed on board of the PCB in dedicated cavities. The ASICS will be in semiconductor packaging and wire bonded to the PCB.
A small sample of FEV11\_COB (same connexion pattern with the interface card than FEV11)
with thickness of 1.2 mm is being produced (see figure \ref{fevcob})
and is planned to be added to the prototype and tested in beam tests conditions. 

\subsection{Data AcQuisition System.}
\label{sec:DAQ}

SKIROC\cite{Callier:2011zz} (Silicon pin Kalorimeter Integrated ReadOut Chip) is the very front end ASIC designed for the readout of the Silicon PIN didoes.
It consists of 64 channels that each comprises a low noise charge preamplifier of variable gain followed by two lines:
a fast line for the trigger decission and a slow line for dual gain charge measurement.
Finally, a Wilkinson type analogue to digital converter fabricates the digitised charge deposition that can be readout. 
Once one channel is triggered, the ASIC reads out all 64 channels adding a bit of information to tag them as triggered or not triggered.
The information is stored in 15 cell deep pyshical switched capacitor array (SCA).
This autotrigger capability is mandatory for ILC case since the accelerator will not provide a central global trigger.
A key feature of the SKIROC ASICs is that they can be power pulsed to meet ILC power consumption requirements.

Every ASU of the SiW-ECAL prototype is equipped with 16 SKIROCs version 2. A new version, 2a, has been produced and will be used to equip new layers currently in production. 

The design of the subsequent chain of the data acquisition (DAQ)\cite{Gastaldi:2014vaa} system is inspired by the ILC.
Current DAQ consists of three modules which are designed to be generic enough to cope with other applications.
The first module is the so called detector interface (DIF) which is placed at the beginning of each layer holding up to 15 ASUs.
All DIFs are connected by single HDMI cables to the concentrator cards: Gigabit Concentrator Card (GDCC).
The HDMI connection is used to transmit both slow control and data readout.
One GDCC controls up to 7 DIFs collecting all data from them and distributing to them the system clock and fast commands.
The most downstream module is the clock and control card (CCC). The CCC provides a clock, control fan-out of up to 8 GDCCs and accepts and distributes external signals (i.e. spill signals).

The whole system is controled by the Calicoes and the Pyrame DAQ software version 3~\cite{pyrame1,pyrame2}. The Pyrame framework provides basic blocks (called modules) of control-command or data acquisition. Calicoes is specific the implementation of these blocks for control-command and data acquisition of the SiW-ECAL prototype. 

\section{SiW-ECAL technological prototype performance on positron beam test.}

This beam test was prepared by a careful and comprehensive commissioning.
The purpose of this phase was to characterize the layers and monitor and control the
noise levels. The commissioning includes the definition of trigger threshold values or noisy channels to be masked.
Studying and control the noise levels in an autotriggered and high-granularity calorimeter
is crucial since noisy cells may saturate the DAQ faster than physical signals.
Two different types of noise were identified:
noise patterns that repeat on all layers 
and noise bursts due to non proper electrical isolation of the layers.
The first source of noise forced us to define a list of channels to be masked in all layers.
The noise bursts issue was circumvented by improving the electrical isolation of single layers.
We have also observed that the noise burst happen only at the end of long acquisitions therefore, in addition to the improved isolation,
we selected short enough acquisitions windows (which indeed are the most appropriate to the high rates of particles in the DESY beam).
However, more studies in the laboratory are needed in order to fully understand these issues.

\begin{figure}[!t]
\centering
\includegraphics[width=3.0in]{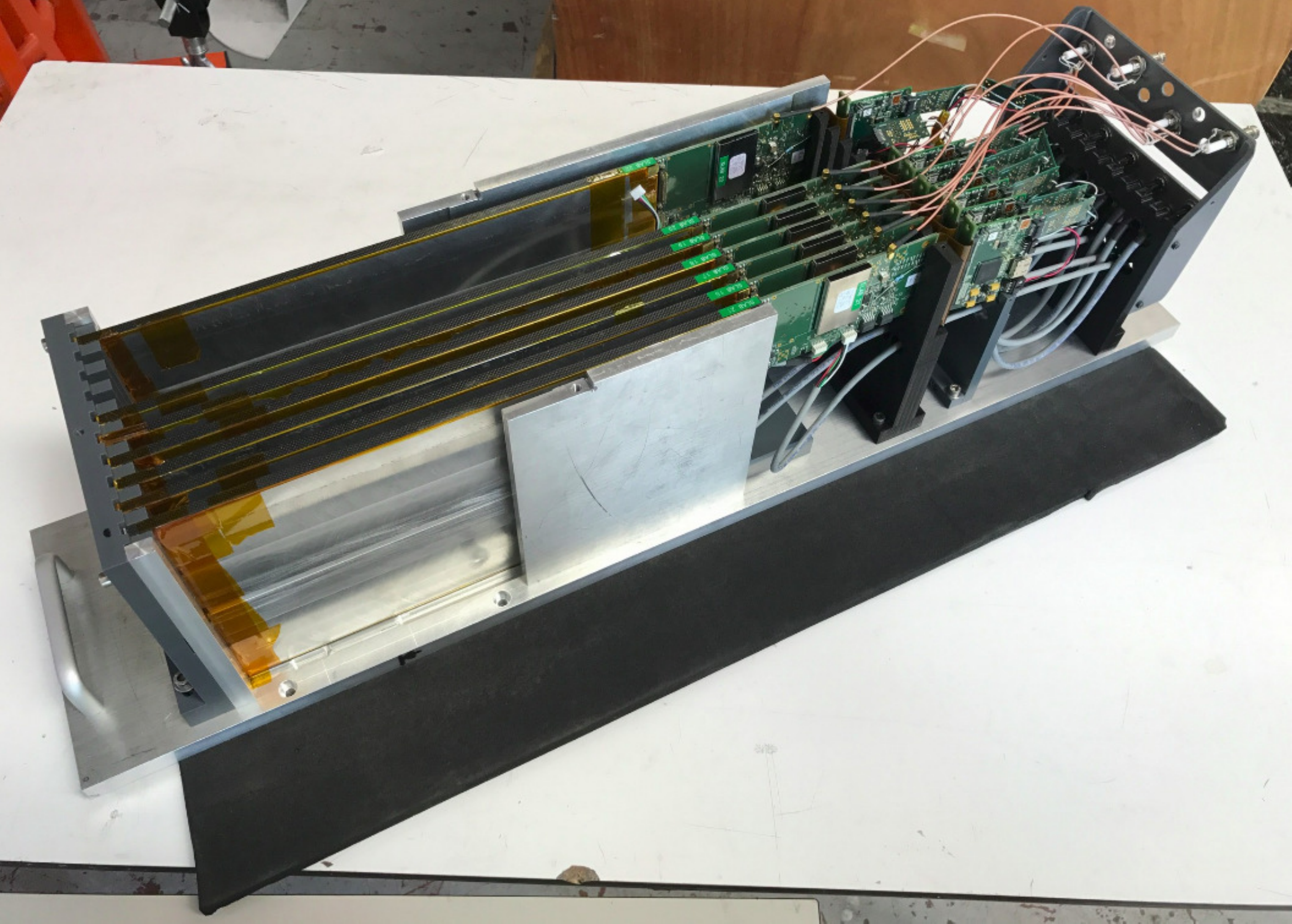}
\caption{Technological prototype: 7 single short slabs inside the mechanical aluminum structure.}
\label{prototype}
\end{figure}

The prototype tested in beam in June 2017 consisted of 7 layers see figure \ref{prototype}.
In the first layer, $\sim 40\%$ of channels were masked due to a damaged Si wafer. In all the other, only the 6-7\% of channels were masked except in the last one were this number grew up to the 16\% due to a faulty ASIC. The number of masked channels may be drastically reduced by setting individual threshold settings instead of global trigger threshold values for each channel on an ASIC. This possibility will become available with the next version of SKIROC ASIC.

The detector was exposed to a positron beam in the DESY test beam area (line 24).
The beam test line at DESY provides continuous positron beams in the energy range of 1 to 6 GeV with
rates of te order of the KHz (with maximum of $\sim 3$ KHz). In addition, DESY gives acces to a bore 1 T solenoid, the PCMAG \cite{pcmag}, in the beam area. 

The detector was running in power pulsing mode without any extra active cooling system.
By means of an external pulse generator we defined the length of the acquisition window to be
3.7 ms at a frequency of 5 Hz. 


The physics program of the beam test can be summarized in the following points:

\begin{itemize}
\item Commissioning and calibration without absorber using 3 GeV positrons acting as minimum ionizing particle (MIPs);
\item magnetic field tests inside the PCMAG magnet with magnetic fields up to 1 T;
\item response to electrons with fully equipped detector, i.e. sensitive parts {\it and} W absorber.
\end{itemize}

\subsection{Calibration runs.}

\begin{figure}[!t]
  \centering
  \begin{tabular}{l}
    \includegraphics[width=2.5in]{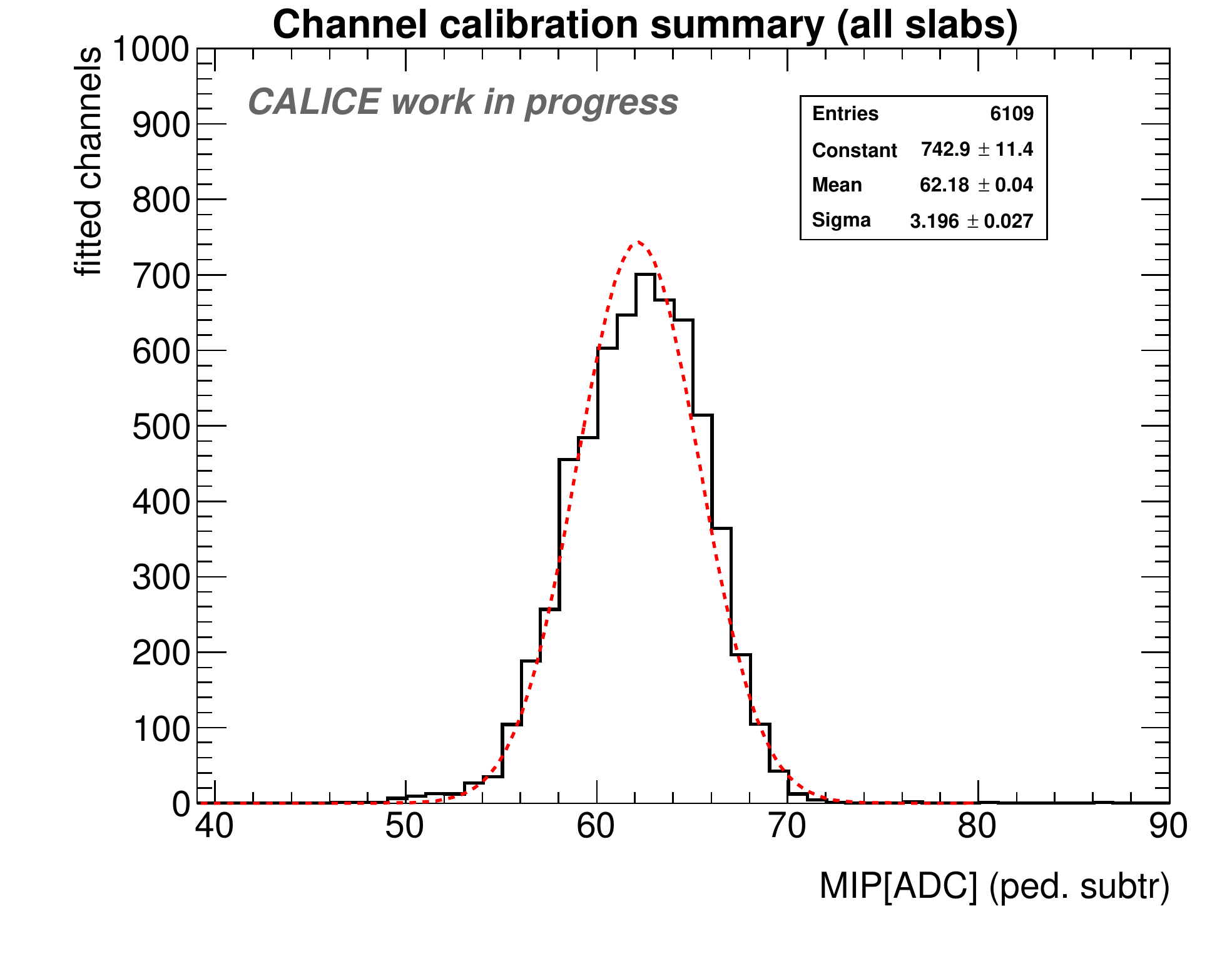}  \\
    \includegraphics[width=2.5in]{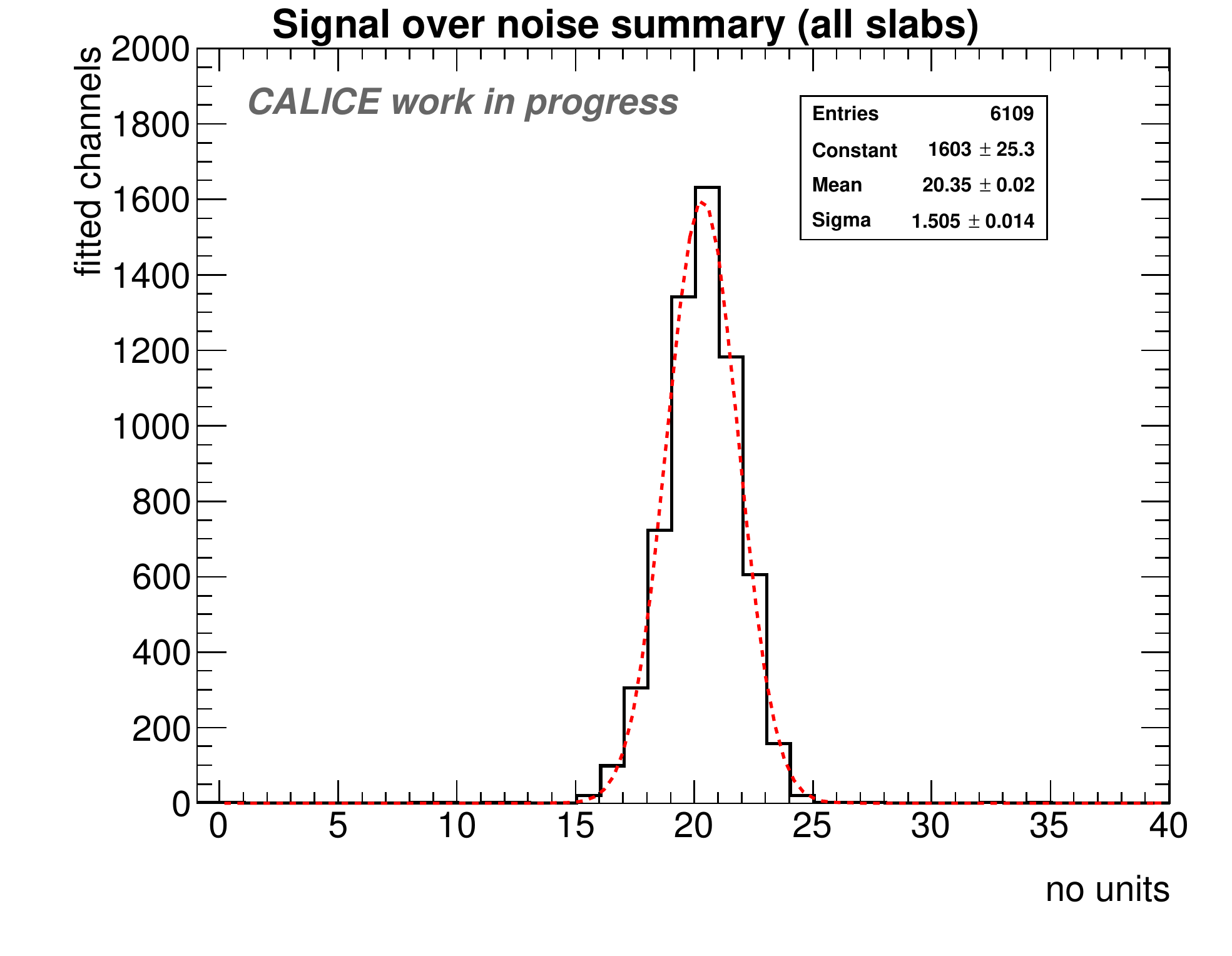}
  \end{tabular}
\caption{Result of the MIP position calculation and signal over noise calculation for all calibrated cells.}
\label{mipandSN}
\end{figure}
  
The main calibration was realized 
by directing the 3 GeV positron beam on 81 positions equally distributed over the surface of the detector.
These data were used for pedestal estimation and energy calibration.
Calibration and pedestal analysis was done for all single layers not requiring requiring track reconstruction.
We calculated the pedestal position for every channel and SCA by fitting the distribution of non triggered hits with a gaussian function.
Afterwards, we subtracted these values to the distribution of triggered hits and fit the resulting distributions to a Landau function convoluted by a gaussian.
The most-probable-value of the convoluted function is taken as the MIP value.
We have obtained a raw energy calibration spread of the 5\% among all cells with the 98\% of all available cells being fitted. Results are summarized in figure \ref{mipandSN}, upper plot.

The signal-over-noise ratio, defined as the ratio between the most-probable-value of
the Landau-gauss function fit to the data (pedestal subtracted) and the pedestal width
(calculated as the standard deviation of the gaussian distribution fitted to the data), was estimated.
The average value for all channels and slabs is 20.3 ($\sim 3$ times better than in the physics prototype, mainly due to the reduced size of cells and different gain values).
Results are summarized in figure \ref{mipandSN}, lower plot.

\begin{figure}[!t]
\centering
\includegraphics[width=3.5in]{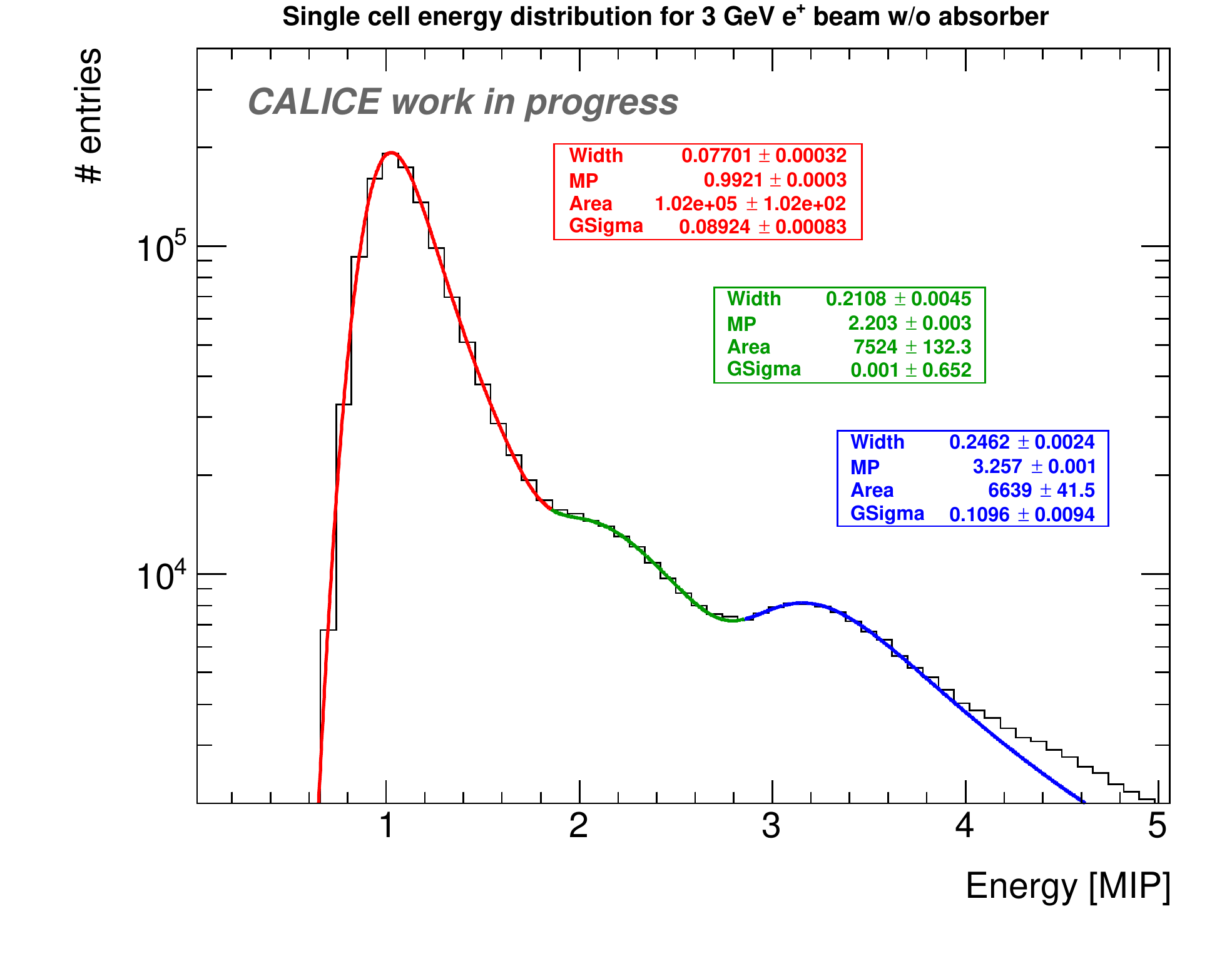}
\caption{Single cell energy distribution (for all calibrated cells) for 3 GeV positron tracks acting as MIPs. }
\label{miplog}
\end{figure}

After pedestal subtraction, calibration and track reconstruction we could finalize the MIP calibration by selecting tracks that cross the detector parallel to its normal.
The result is shown in figure \ref{miplog} where single cell the energy distribution for MIPs is shown for all calibrated cells.
The distribution reveals the presence of a second and third peak due to events of multiparticles crossing the detector.

Finally, a calibration run with the beam hitting the slabs under an angle of $\sim 45^{0}$ was done.
The purpose of this run was to proof that the MIP position scales with a factor $\sqrt{2}$
due to the larger path to be crossed by the electrons in the Si wafer.
Preliminary results show a perfect agreement with the expected results.

\subsection{Magnetic field tests.}

For this test, a special PVC structure was
designed and produced to hold the slab vertically, perpendicular to the beam.

The purpose of the test was twofold: first to proof that the DAQ, all electronic devices and the mechanical slab itself were able
to handle strong magnetic fields; second purpose was to proof the stability of performance during these tests.
We took several runs, with 0, 0.5 and 1 T magnetic fields with and without 3 GeV positron beam.
We observed that the pedestals position is independent of the magnetic field (within the 1 per mile). The MIP position was increased, in average,
of the 3\% for 1 T and 1.5\% for 0.5 T with respect with the 0 T case.
This level of increase is expected since the positron traversing the magnetic field hit the slab with a deflecting angle,
increasing then the path of the particle through the wafer. More detailed studies and simulation comparisons are to come.

\subsection{Response to electromagnetic showers.}

\begin{figure}[!t]
  \centering
  \begin{tabular}{l}
    \includegraphics[width=3.0in]{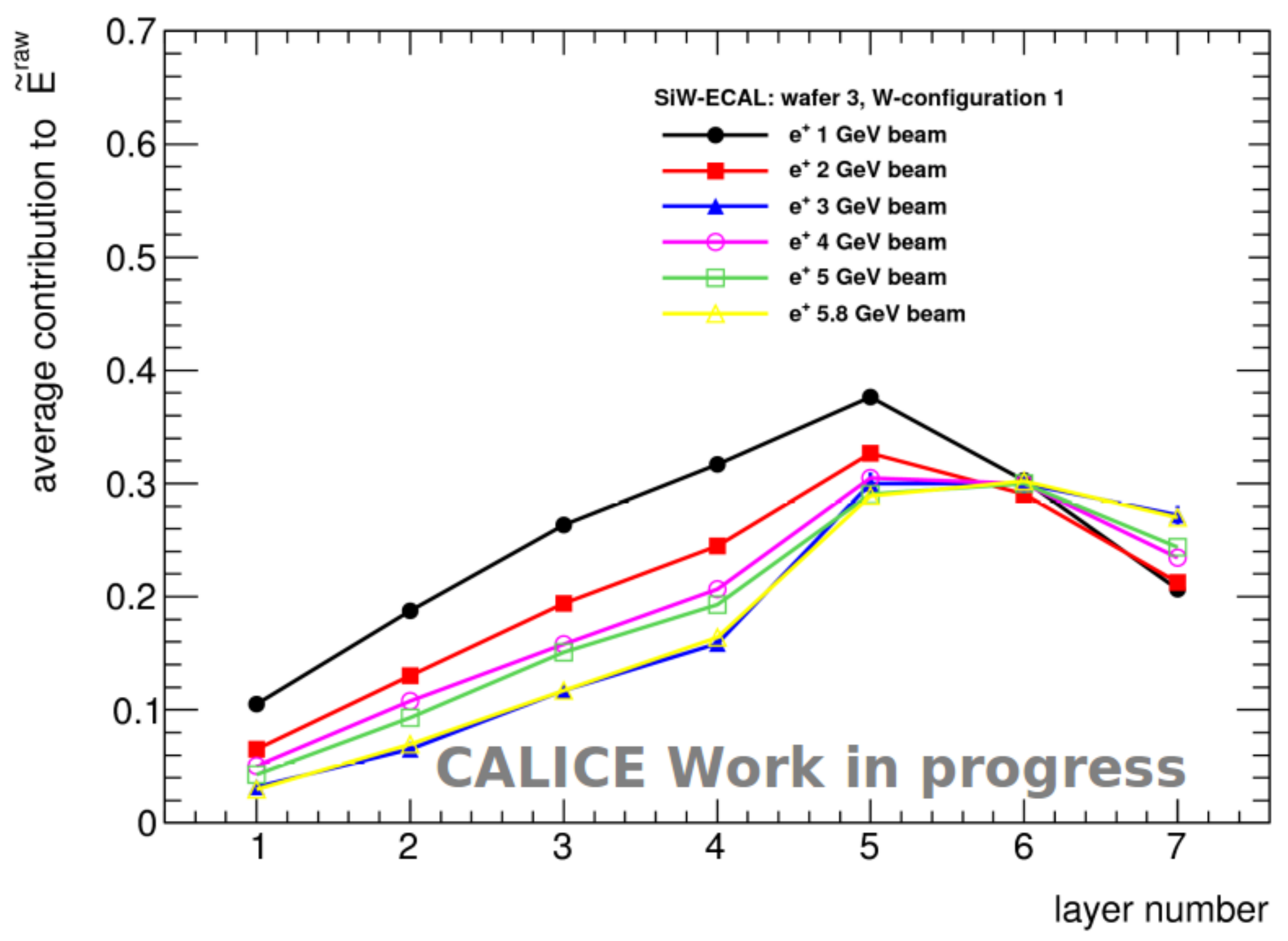}\\
    \includegraphics[width=3.0in]{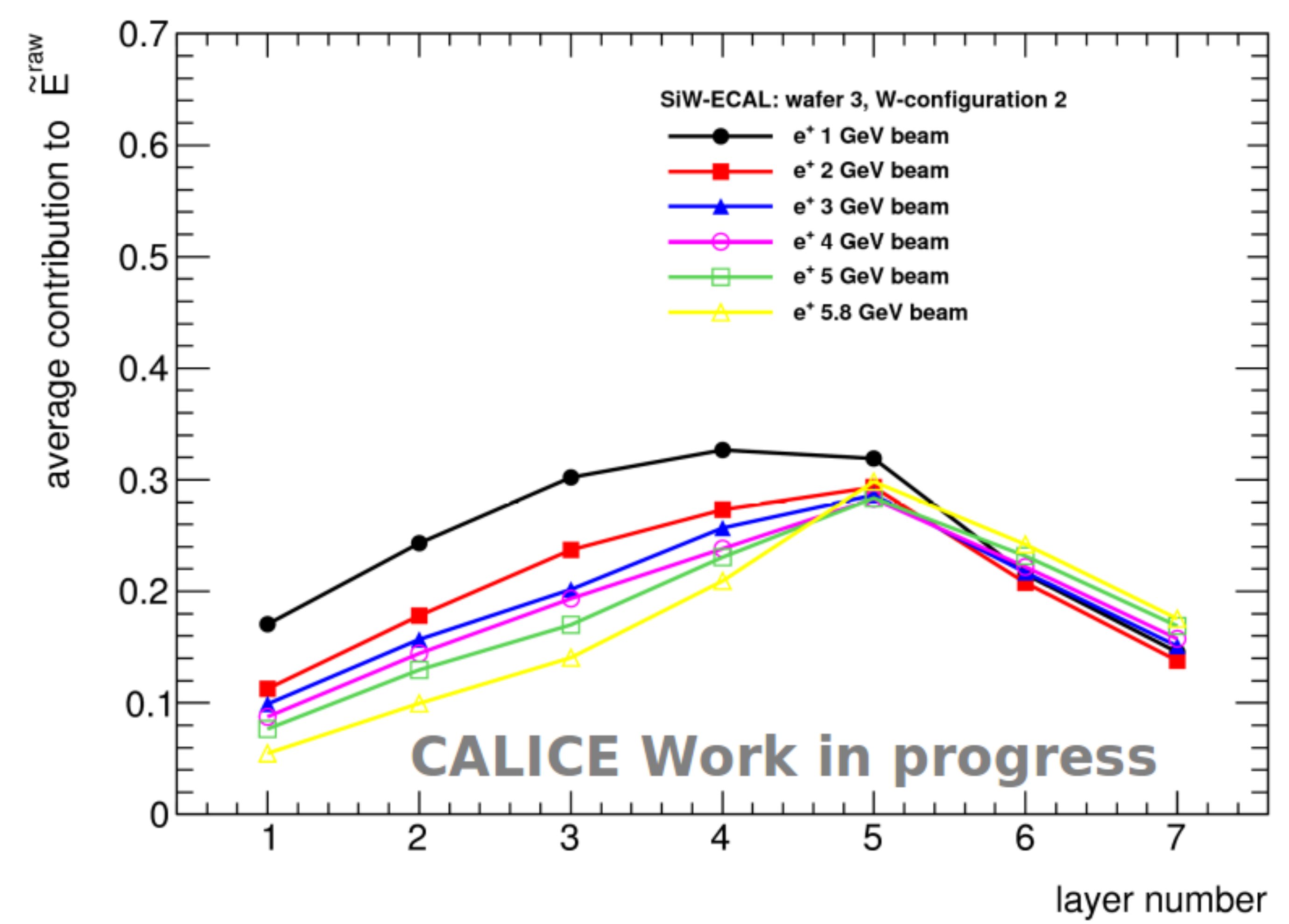}\\
    \includegraphics[width=3.0in]{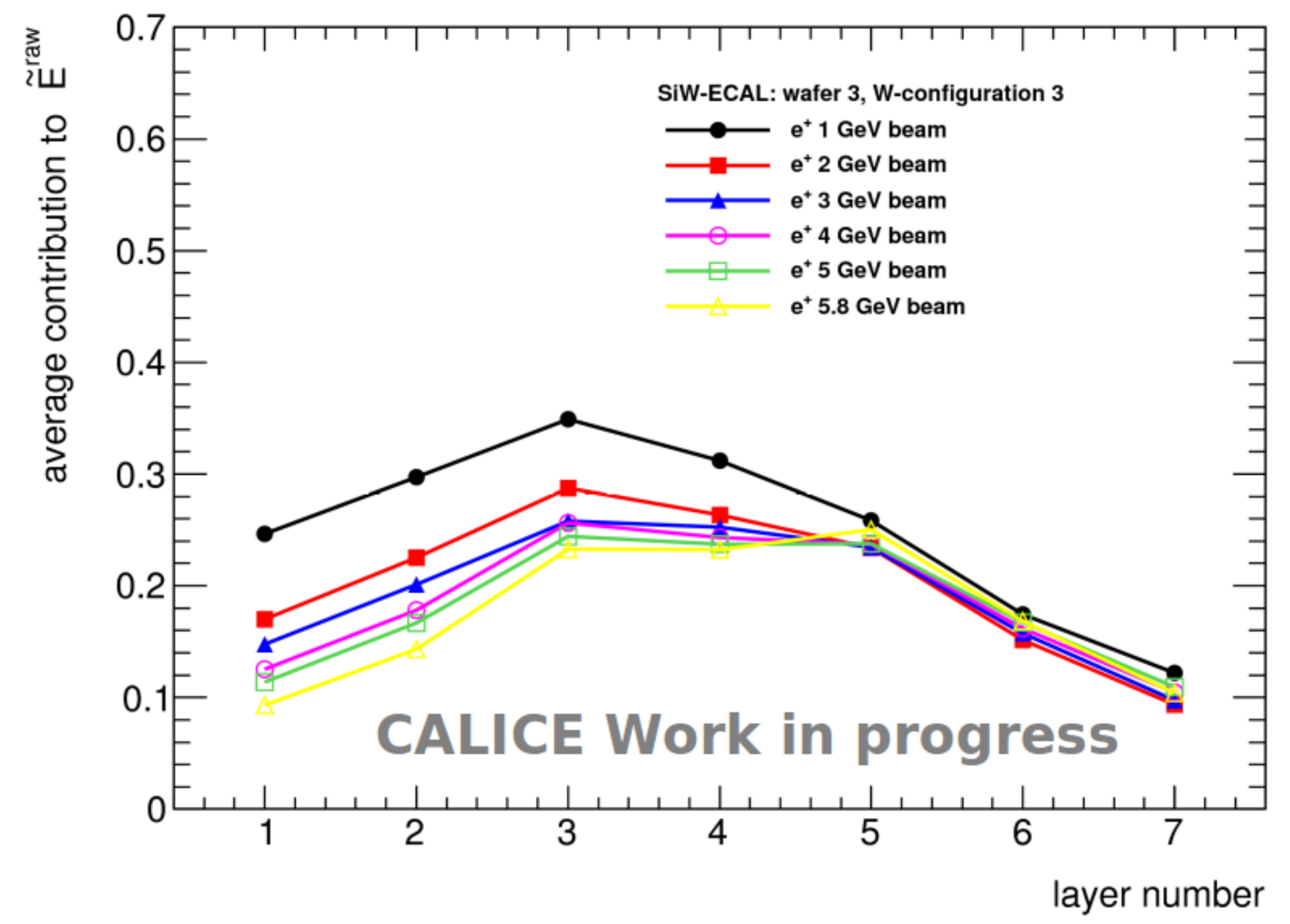}
  \end{tabular}
\caption{Raw electromagnetic shower profiles for three different tungsten configurations. In the x-axis, we show the layer number. In the y-axis, the averaged fraction of {\it energy} (sum of ADC in all triggered cells in a event, considering only events where all layers had at least a hit) measured in every layer. }
\label{showers}
\end{figure}

The purpose of the beam test was to study the interaction of electrons with the absorber material resulting in
electromagnetic showers.
We inserted W plates of different thicknesses between the sensitive layers and we performed
a scan of energies of the positron beam: 1, 2, 3, 4, 5 and 5.8 GeV.
We tested the response of the detector with three different configurations of the W repartition.
The first one adding up a total of 6.16 $X_{0}$, the second 7.84 $X_{0}$ and the third of 9.52 $X_{0}$.
Preliminary results of the raw electromagnetic shower profiles are shown in figure \ref{showers}
for the three different configurations and for all energies. This first approach to the data looks promising but further studies and comparisons with simulations are needed.

\section{Outlook.}

\begin{figure}[!t]
\centering
\includegraphics[width=2.0in]{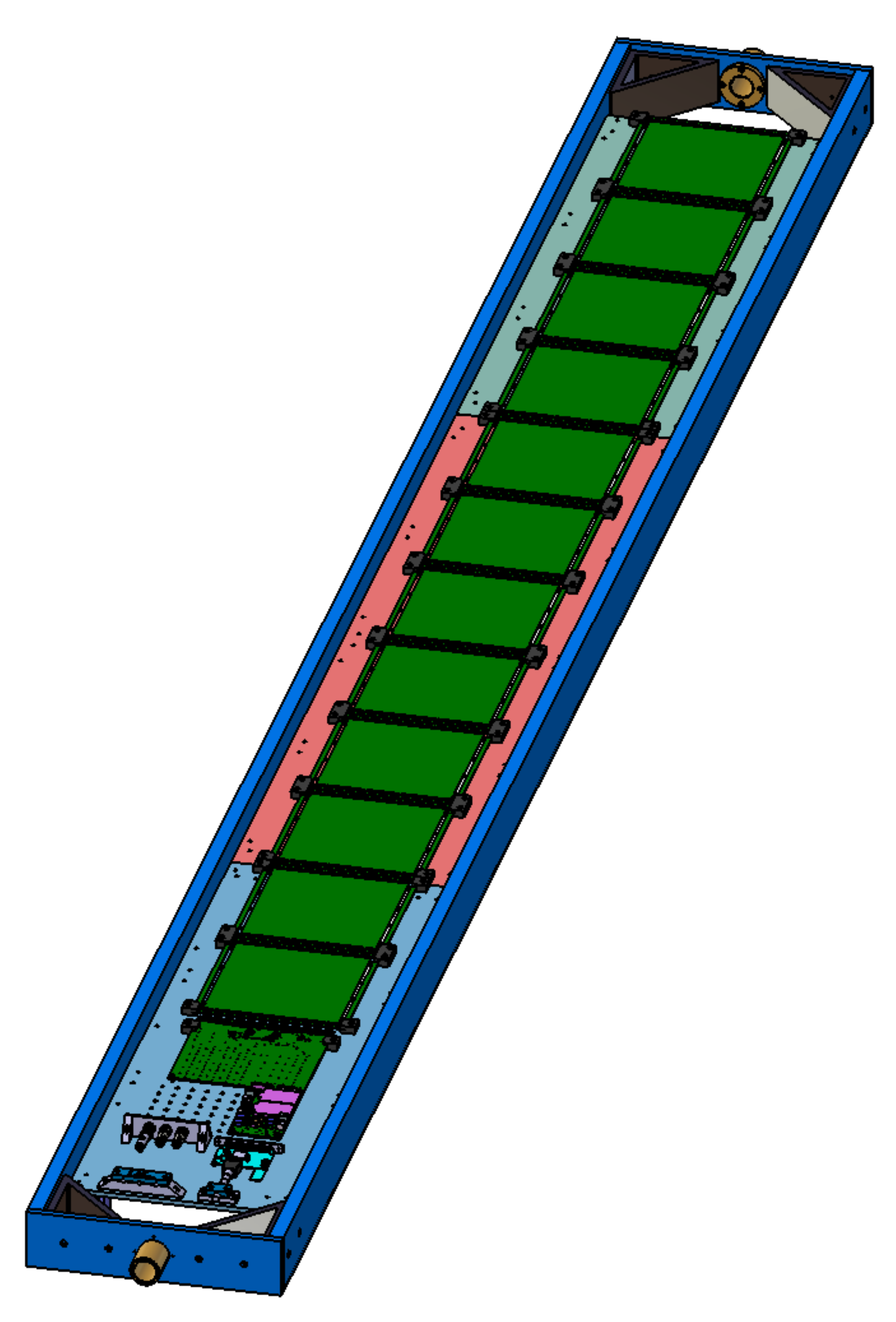}
\caption{Design of a long layer being currently produced.}
\label{longslab}
\end{figure}

The ILD ECAL will host long layers of up to $\sim$2.5m.
Current research efforts are focused in the construction and test of such long layers made of chains of ASUs (up to $\sim$15 ASU).
A long layer of $\sim$2.m  is now being produced and tested in the laboratory (figure \ref{longslab}).
A long layer constitutes a technological challenge in both aspects, the mechanical
(very thin and long structure with fragile sensors in the bottom, complicated assembly procedure...)
and the electrical (i.e. transmission of signals and high currents).
For example, interconnections between ASUs and between ASU and interface card are one of the most involved parts of the assembly
and require close collaboration between mechanical and electronic engineers.
Currently, this interconnection is made with flat kapton connectors manually soldiered onto the layers by heating procedure.
This is, so far, carried manually and found to be time consuming. Other alternatives are currently being  investigated.

\appendices
\section*{Acknowledgment}

This project has received funding from the European Union’s Horizon 2020 Research and Innovation programme under Grant Agreement no. 654168.

This work was supported by the P2IO LabEx (ANR-10-LABX-0038), excellence project HIGHTEC,
in the framework {\textquotesingle}Investissements d{\textquotesingle}Avenir{\textquotesingle}
(ANR-11-IDEX-0003-01) managed by the French National Research Agency (ANR).

The research leading to these results has received funding from the People Programme (Marie
Curie Actions) of the European Union{\textquotesingle}s Seventh Framework Programme (FP7/2007-2013)
under REA grant agreement, PCOFUND-GA-2013-609102, through the PRESTIGE
programme coordinated by Campus France.

The measurements leading to these results have been performed at the Test Beam Facility at DESY Hamburg (Germany), a member of the Helmholtz Association (HGF).

\ifCLASSOPTIONcaptionsoff
  \newpage
\fi

%








\end{document}